\documentclass{PoS}

\title{Summary of Working Group 5: Physics with Heavy Flavours\\ $\phantom{-}$ \\
\normalsize \rm TUM-HEP-1221/19}

\ShortTitle{Summary of Working Group 5: Heavy Flavours}

\author{V. Bhardwaj\\
        Indian Institute of Science Education and Research Mohali\\
        E-mail: \email{vishstar@gmail.com}}
\author{J. Libby\\
        Indian Institute of Technology Madras\\
        E-mail: \email{libby@iitm.ac.in}}
\author{J. Virto \\
Center for Theoretical Physics, Massachusetts Institute of Technology\\
Physics Department T31, Technische Universit\"{a}t M\"{u}nchen \\
E-mail: \email{jvirto@mit.edu}}

\abstract{We present a summary of heavy-flavour physics at the the time of the DIS2019 conference. This summary is based upon the relevant plenary and parallel talks presented at the workshop. The summary is divided into four broad areas: (1) spectroscopy, (2) the production of $b$ and $c$ quarks in collisions, (3) top quark production and properties, and (4) $b$ and $c$ quark decay.}

\FullConference{XXVII International Workshop on Deep-Inelastic Scattering and Related Subjects - DIS2019\\
		8-12 April, 2019\\
		Torino, Italy}

\begin{document}

\section{Introduction}
The heavy flavour working group at DIS2019 is very much one of the related subjects referred to in the full title of this workshop. The relationship to deep-inelastic scattering (DIS) differs depending on the topic under consideration. Understanding the nature of bound states of quark matter is one of the key facets of DIS. A plethora of states containing heavy quarks have been discovered since the $e^{+}e^{-}$ $B$ factories began operation in 1999. Many of these states are considered exotic due to the fact that they do not appear to be baryons or mesons; these discoveries of exotic states have continued with LHCb and BESIII. Therefore, the first topic we will review, in Sec.~\ref{sec:spectro}, is spectroscopy. Next, in Sec.~\ref{sec:prod}, we will turn to the production of $b$ and $c$ quarks in collisions, particularly proton-proton, proton-nuclei and nuclei-nuclei, where much of the focus is now on what can be learned about the quark-gluon plasma (QGP) from these interactions. The top-quark is special in that its natural width is very much greater than the scale of quantum chromodynamics (QCD), which results in it decaying before it can hadronise. Therefore, its production and decay can shed light on both electroweak and QCD aspects of the standard model (SM); Section~\ref{sec:top} is dedicated to a review of recent top-quark physics results. Finally, we review some results related to the decay of $b$ and $c$ quarks in Sec.~\ref{sec:decay}; several of these measurements may be shedding some light on physics outside the SM, the quest for which is a goal of the DIS community. 

Before we begin these summaries, we would like to note the one result presented that is not a related subject, but actually a DIS measurement. The ZEUS Collaboration presented a measurement of the inclusive $ep\to b\overline{b} X$ cross section with the full HERA data sample corresponding to an integrated luminosity of $380~\mathrm{pb}^{-1}$ \cite{bib:zeus_bbinclusive}. The measured value is $11.4\pm 0.8^{+3.9}_{-2.8}~\mathrm{nb}$.  Here, and elsewhere in these proceedings, if two uncertainties are quoted, the first uncertainty is statistical and the second is systematic. The measurement is in agreement with the predicted value of $7.5^{+4.5}_{-2.1}$~nb from next-to-leading order (NLO) calculations \cite{bib:bbinclusive_theory}. The study also included differential cross section measurements for the first time. To have such measurements appearing, even 12 years after HERA was decommissioned, is a testament to the unique nature of the data set collected. 
\section{Spectroscopy}
\label{sec:spectro}
Spectroscopy with heavy quarks is an excellent place to test QCD, particularly in relation to exotic forms of quark matter. The discovery of the so-called $XYZ$ states, which do not fit the expected pattern of pure $c\overline{c}$ or $b\overline{b}$ bound states, have led to alternate theories to describe their composition. (For a recent review of the $XYZ$ states see Ref.~\cite{xyz_review}.) The question of whether these states are tetraquarks (tightly bound states of two quarks and two antiquarks), di-meson molecules, or some hybrid state of mesons and gluons has remained unresolved.  
The BESIII Collaboration has been studying the relative branching fractions of the $X(3872)$ with respect to $X(3872)\to J/\psi \pi^{+}\pi^{-}$ decay. The first observation of $X(3872)\to \pi^0\chi_{c1}$ decays \cite{besIII_Xtochic1pi0} may be indicative of a tetraquark composition. The observation of $X(3872)\to \omega J/\psi$ \cite{besIII_XtoomegaJpsi} confirms the significant isospin violation observed in the decays of the $X(3872)$. The BESIII Collaboration also reported the first observation of the $Y(4220)$ decaying to open charm \cite{bes3_YtoDDpi}, in a manner consistent with the interpretation of the $Y(4220)$ as a $D\overline{D}_1(2420)$ molecule \cite{DDmolecule}. There are two recent studies of the charged $Z$ states by LHCb Collaboration. The first study is an angular analysis of the decay $B^0\to J/\psi K^+\pi^-$ to determine whether the $Z_c^-$-like structures in the $J/\psi\pi^-$ invariant-mass spectrum are the result of high-mass $K^{*}$ resonances; the $K^{*}$ only hypothesis is excluded by more than $10\sigma$ \cite{lhcb_B0toJpsiKpi}, where $\sigma$ is a standard deviation, and the exotic structures confirmed in a model-independent manner. The second study provides the first evidence of a $Z_c^-$ structure in the decay $B^{0}\to \eta_c(1S)\pi^-K^+$ \cite{lhcb_B0toetacKpi} with a mass and width of $4096\pm20^{+18}_{-22}$~MeV and $152\pm58^{+60}_{-35}$~MeV, respectively. Despite the huge amount of information that has been accumulated about the $XYZ$ states, their exact nature will only be elucidated with the analysis of additional data. Fortunately, the prospects are good with BESIII performing a scan of the $e^+e^-$ invariant mass interval between 4.2 and 4.6 GeV, more data to be analysed and collected by LHCb, and the start of Belle II \cite{b2tip}.  

\begin{figure}[t]
\begin{center}
 \includegraphics[width=0.45\columnwidth]{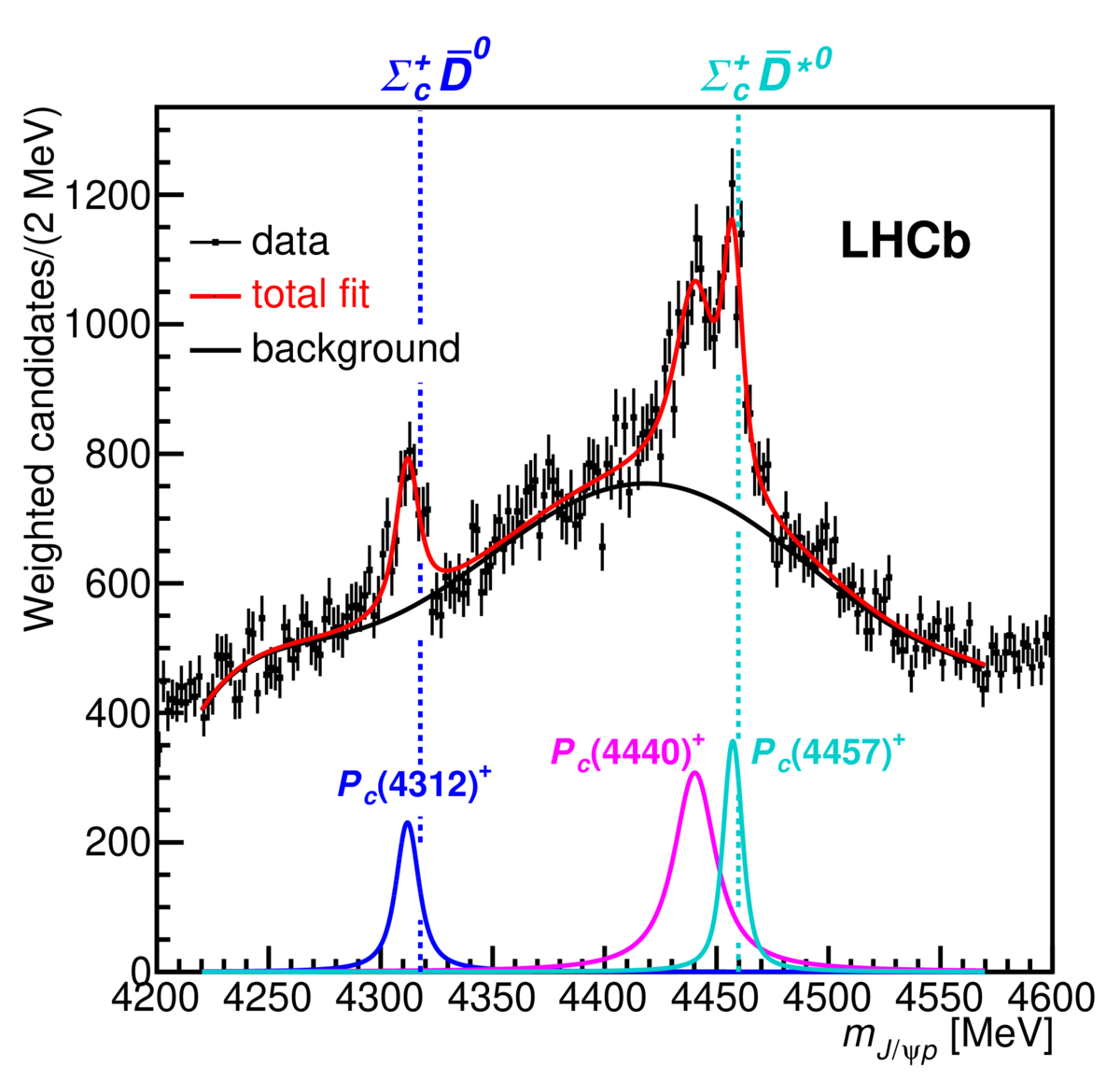}
\caption{Invariant mass spectrum of $J/\psi p$ from $\Lambda_b^0\to J/\psi p K^-$ decays measured be LHCb \cite{lhcb_pentaquark2019}. Fits to the three pentaquark states are shown along with the  $\Sigma_c^{+}\overline{D}^{(*)0}$ thresholds.}\label{fig:lhcb_pentaquarks}
\end{center}
\end{figure}

The outstanding result on spectroscopy in 2019 has been the observation of the narrow states in the $J/\psi p$ invariant mass spectrum from $\Lambda_b^0\to J/\psi p K^-$ decays. Previously, LHCb had observed that the $J/\psi p$ invariant mass spectrum from these decays could only be well described when a broad $P_c(4380)^+$ state and a narrow $P_c(4450)^{+}$ state are included \cite{lhcb_pentaquark2015}. These new states are interpreted as $c\bar{c}uud$ pentaquarks. A new analysis has been performed with a sample of  $\Lambda_b^0\to J/\psi p K^-$ decays that is an order of magnitude larger than that previously analysed \cite{lhcb_pentaquark2019}. The analysis identifies a new narrow state, the $P_c(4312)^+$, and that the originally observed $P_c(4450)^+$ is  composed of two overlapping narrow states the $P_c(4440)^+$ and $P_c(4457)^+$. Figure~\ref{fig:lhcb_pentaquarks} shows a fit to the $J/\psi p$ invariant-mass spectrum including these states. The masses of the states lie just below the sum of the $\Sigma_c^{+}$ and $\overline{D}^{(*)0}$ masses, which is strongly indicative of a molecular explanation for these states. However, to confirm the molecular interpretation, both a full partial-wave analysis to determine the spin and parity of the states, as well as observation of isospin partners, are required; as such some contribution to these states from a tightly bound five-quark state cannot be ruled out. 
 
\section{$b$ and $c$ quark production}
\label{sec:prod}
Before discussing the $b$ and $c$ quark production, we will highlight an important measurement involving the production of strange baryons, specifically the determination of the $\Lambda$ polarisation by the BESIII Collaboration \cite{lambda_polarization}. Exploiting a sample of $1.3$ billion $J/\psi$ mesons, a full angular analysis of the process $e^{+}e^{-}\to J/\psi \to \Lambda\bar{\Lambda}$ allows the polarization to be determined for both $\Lambda$ and $\bar{\Lambda}$ separately. The previous measurements of the polarization are found to disagree with the new measurement by $(-17\pm3)\%$, which means that all measurements that previously relied on these parameters should be reinterpreted. Further, the difference between the polarisation parameters for $\Lambda$ and $\bar{\Lambda}$ provide a $1\%$-level test of $CP$ violation in hyperon decays.

Many results for quarkonium production are available from RHIC and the LHC. The differing centre-of-mass energy and pseudorapidity coverage of the experiments lead to complementarity among the experiments. The initial states studied are: $(i)$ $pp$, as a reference for $pA$ and $AA$, as well to better model production for $pp$ analyses; $(ii)$ $pA$ as a means to investigate the transition between cold to hot nuclear matter {\it i.e.} the effect of nuclear shadowing; and $(iii)$ $AA$ to study hot nuclear matter and search for evidence of the QGP. The suppression of $c\bar{c}$ and $b\bar{b}$ state production has been considered a key signature of the QGP \cite{onia_qgp}. There are many observations of suppression of charmonium and bottomonium in $AA$ collisions since the first suggestions from the NA50 experiment \cite{NA50}. Recent results have focused on testing models of onia production and understanding what, if any, impact cold nuclear matter (CNM) effects have. For example, there are several measurements comparing $p\mathrm{Pb}$ and PbPb production rates \cite{ATLAS_jpsi,CMS_jpsi,ALICE_jpsi}, which confirm the suppression is unlikely to be the result of CNM effects. Further, sequential suppression \cite{sequential_suppression} between the $J/\psi$ and $\psi(2S)$ is observed. Precise tests of the QGP using quarkonia are expected in the years ahead that will hopefully allow a consistent model to be developed to describe all the measurements.

\section{Top quark physics}
\label{sec:top}
The rate of $t\bar{t}$ pair production at the LHC is 8~Hz, when the collider is operating at an instantaneous luminosity of $10^{34}~\mathrm{cm}^{-2}\mathrm{s}^{-1}$ and a $pp$ centre-of-mass of energy of $13~\mathrm{TeV}$. Therefore, the LHC is a {\it top factory} that is accumulating unprecedented samples to study top production (Sec.~\ref{subsec:top_production}) and the top quark properties, such as its mass, width, spin correlations and couplings (Sec.~\ref{subsec:top_properties}). The wide array of measurements provide information across the electroweak and QCD sectors of the SM, as well as probing for beyond-the-SM contributions.    

\begin{figure}[t]
\begin{center}
 \includegraphics[width=\columnwidth]{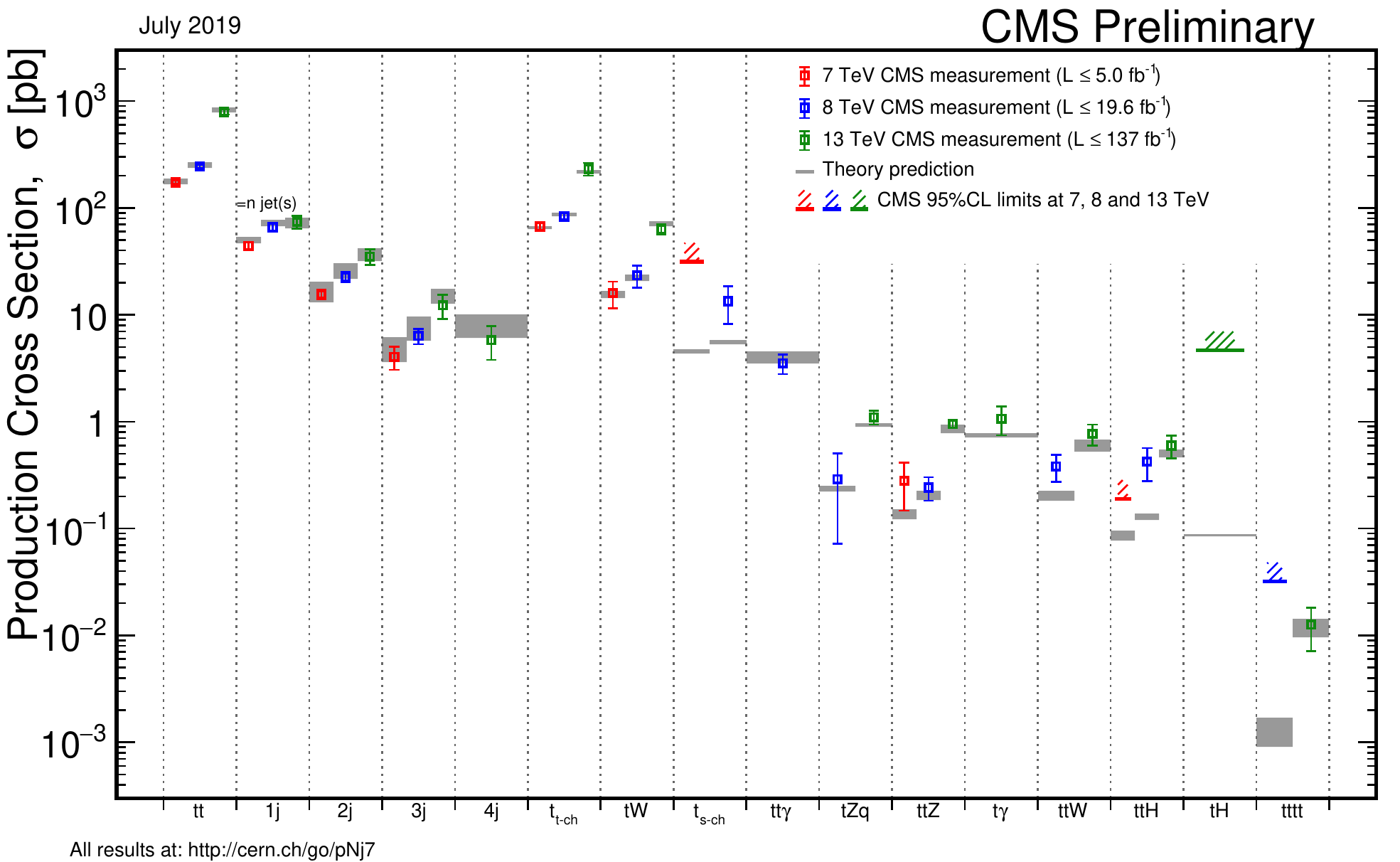}
\caption{Summary of top pair, single-top and four top production cross section measurements and limits presented by the CMS Collaboration. The measurements include those of top production associated with other particles. Standard model predictions are shown as grey bands for comparison.}\label{fig:top_xsec}
\end{center}
\end{figure}

\subsection{Top production measurements}
\label{subsec:top_production}
Figure~\ref{fig:top_xsec} summarises the many measurements of top cross sections for pair production and single top production reported by the CMS Collaboration, including measurements where additional particles are produced in association.\footnote{Similar plots compiled by the ATLAS Collaboration can be found at \href{\tt https://atlas.web.cern.ch/Atlas/GROUPS/PHYSICS/CombinedSummaryPlots/}{https://atlas.web.cern.ch/Atlas/GROUPS/PHYSICS/CombinedSummaryPlots/}.} 
These measurements cover cross sections that span five orders of magnitude, thus allowing precision tests of the theoretical predictions even of relatively rare processes. 
We cannot cover all of these measurements, so we will focus on some recent developments related to production of top-pairs and single-top in association with a vector boson. 

The CMS Collaboration  has recently reported measurements of $pp\to t\bar{t}Z^{0}$ production  \cite{CMS:2019too} using a $13~\mathrm{TeV}$ data sample corresponding to 77.5~fb$^{-1}$. This process is directly sensitive to the $Z^{0}\to t\bar{t}$ coupling. These measurements use decays that have three or four charged leptons in the final state, which include an oppositely charged pair consistent with $Z^0\to \ell^+\ell^-$ decay, as well as $b$-tagged jets from the top-quark decay. The production cross section is measured to be $\sigma(t\bar{t}Z)=0.95\pm0.05\pm 0.06$~pb. The samples are of sufficient size to allow differential cross section measurements for the first time. These differential distributions are fit in the context of SM effective-field theory (SMEFT) \cite{warsaw,smeft_interpretation} to obtain the most stringent limits to date on anomalous couplings of the top to the $Z^{0}$ boson. Such a SMEFT interpretation has great potential to be applied to other measurements of associated production such as $t\bar{t}W$ and $t\bar{q}W$. 

The ATLAS Collaboration has recently made measurements of the $tW$ cross section using a sample of 13~TeV collision data, corresponding to an integrated luminosity of 36.1~fb$^{-1}$ \cite{atlas_singletop}. The final state is identified by the presence of two charged leptons and two $b$-tagged jets. The differential cross section with respect to the $bl^{+}$ mass is measured and to fully describe the data away from the peak it is observed that the interference with the $gg\to t\bar{t}\to b\bar{b}\ell^{-}\bar{\nu}_{\ell}\ell^{+}\nu_{\ell}$ process must be accounted for. Recent NLO calculations at the parton level \cite{bb4l} have made this possible, further the narrow top width approximation has to be relaxed to include the interference, which can in turn be used to determine the top width, as we will describe in next section.  

\subsection{Top properties}
\label{subsec:top_properties}
Given the top mass is known to 0.2\% \cite{pdg} much more attention is now paid to determining its other properties such as width, spin correlations and couplings. The differential $tW$ cross section \cite{atlas_singletop} already discussed has allowed a direct determination of the top width \cite{top_width} by using the interference in the tails of the $b\ell$ invariant mass distribution. The value obtained is $1.28\pm 0.30$~GeV, which agrees with the SM prediction of $1.33\pm 0.29$~GeV, and does not require the restrictive assumption of the indirect determinations \cite{width_indirect} that the branching fraction for $t\to bW^+$ is 100\%.  

\begin{figure}[t]
\begin{center}
 \includegraphics[height=0.45\columnwidth]{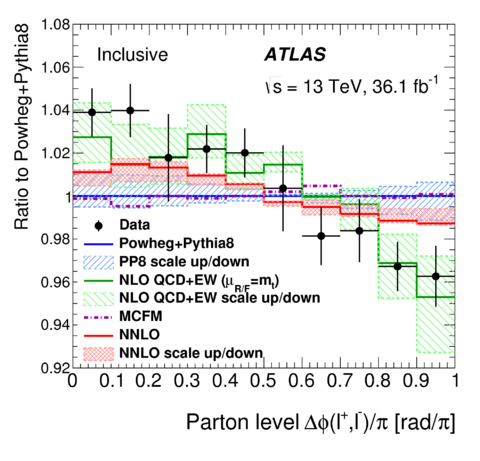}
\caption{Relative deviation of $\Delta\phi$ in $t\bar{t}$ events measured by the ATLAS Collaboration \cite{spin_correlations} to the generator prediction from {\tt POWHEG + PYTHIA8}. Several calculations, along with their uncertainties are shown, with only the fixed-order calculation shown in green giving reasonable agreement.}\label{fig:top_corr}
\end{center}
\end{figure}

 The spin correlations of the produced $t\bar{t}$ pair at the LHC are preserved due to the top lifetime being very much less than the hadronisation timescale. Information about the correlation is accessed in events where both $W$ boson daughters of the $t\bar{t}$ pair decay leptonically, by measuring the azimuthal opening angle $\Delta\phi$ between the two leptons. The ATLAS Collaboration has reported results using 13~TeV $pp$ collision data corresponding to an integrated luminosity of 36.1~fb$^{-1}$ \cite{spin_correlations}. The distribution of $\Delta\phi$ is compared to several NLO generator predictions, which consistently underestimate the size of the correlation (Fig.~\ref{fig:top_corr}). However, a fixed order calculation \cite{corr_calculation1,corr_calculation2,corr_calculation3} agrees with the data within the large uncertainties related to the assumed scales in the calculation. Therefore, more experimental and theoretical work is required to assess whether the observed enhancement maybe the result of possible beyond-the-SM physics.

The final top-physics topic we discuss is a proposal to measure the magnitude of the Cabibbo-Kobayashi-Maskawa matrix element $V_{cb}$ using top decays \cite{top_Vcb}. This would be the first measurement at the electroweak scale rather than that of the $b$-quark mass. All previous measurements of $V_{cb}$ use semileptonic $b$ decays, which have a systematically limited precision of 2\% \cite{hflav} due to the theoretical interpretation of the measurements. Furthermore, there has been long-standing tension between measurements made in exclusive and inclusive $b$ decays, so alternative methods are desirable. The proposed method uses $t\bar{t}$ pairs where $\bar{t}\to\bar{b}W^{-}\to\bar{b}\ell^{-}\bar{\nu}_{\ell}$ and $t\to bW^{+}\to b\bar{q}c$. The fraction of events with $q=b$ will be proportional to $\left|V_{cb}\right|^2$. Therefore, the experimental topology is a lepton, a charm-tagged jet and three $b$-tagged jets. It is estimated  that with a HL-LHC data set of 3000~fb$^{-1}$ a relative precision of 1 to 5\% on $|V_{cb}|$ is possible, where the variation arises from assumptions about the performance of the charm-tagging algorithm \cite{HL_Vcb}.

\section{$b$ and $c$ decays}
\label{sec:decay}
Our final topic is the study of weak decays of $b$ and $c$ quarks, otherwise known as flavour physics. The current focus in flavour physics is to make precision measurements of observables that are potentially sensitive to physics beyond the SM. We review some recent measurements of $CP$ violation and tests of lepton universality. We also consider the interpretation of the observed anomalies in $B$ decays.

$CP$ violation has been observed in the decays of $s$ and $b$ quarks for some time, further, the amount of $CP$ violation observed is described well by the Kobayashi-Maskawa mechanism \cite{KM} in the SM. However, $CP$ violation had not been observed in `up'-type quarks. $CP$ violation is expected to be very small in the SM but is most likely to be enhanced in singly Cabibbo suppressed (SCS) decays \cite{SCS}. The LHCb Collaboration reported the first observation of $CP$ violation in charm meson decay this year \cite{LHCb_charmCPV}. The LHCb Collaboration measures $\Delta A_{CP}$, which is the difference in direct $CP$ asymmetries measured separately in the SCS decays $D^0\to \pi^+\pi^-$ and $D^{0}\to K^{+}K^{-}$, because it is experimentally robust against asymmetries in the production and detection of the final state. Combining all the data collected at 7, 8 and 13~TeV, as well as tagging the flavour of the $D^0$ using the decay chains $D^{*+}\to D^{0}\pi^+$ and $\overline{B}\to D^0\mu^-\overline{\nu}_{\mu}X$, they measure $\Delta A_{CP} = \left(-15.4\pm 2.9\right)
\times 10^{-4}$, which differs from zero by more than $5\sigma$. Theoretical predictions based on light-cone sum rules \cite{firstprinciple_QCD} lead to upper bounds for $|\Delta A_{CP}|$ which are roughly an order of magnitude smaller than the experimental central value. Howerver, the large asymetry is so far consistent with SU(3)-flavor analyses (see for example Ref.\cite{sum_rules}).
Only further measurements of $CP$ violation in the charm system, along with theoretical advances, will clarify whether new physics is being observed in charm $CP$ violation.

The other recent experimental advances have come from additional measurements of modes in which the so called flavour anomalies have been observed: $B\to K^{\left(*\right)}\ell^+\ell^-$ and $B\to D^{\left(*\right)}\tau^{-}\overline{\nu}_{\tau}$. Two of the anomalies are related to possible violations of lepton universality. The first are measurements of $R_{K^{(\ast)}}=\frac{\mathcal{B}\left(B\to K^{(\ast)}\mu^{+}\mu^{-}\right)}{\mathcal{B}\left(B\to K^{(\ast)}e^{+}e^{-}\right)}$, which have been observed to be consistently lower than the SM prediction by between $2\sigma$ to $3\sigma$ in different regions of $q^2$, the dilepton invariant mass, by the LHCb Collaboration \cite{RK_Run1,RKstar_Run1}.
These processes are electroweak-penguin decays, whose leading-order amplitudes proceed at loop level, hence, these observables are very sensitive to possible contributions from new physics.
Until this year, measurements were only reported using data samples collected at centre-of-mass energies of 7 and 8 TeV, which corresponded to an integrated luminosity of 3~$\mathrm{fb}^{-1}$. An updated measurement of $R_{K}$ has been released with an additional 2~fb$^{-1}$ of data included, which was accumulated at 13~TeV \cite{RK_update}. The measured value is $R_{K}\left(1.1<q^2<6.0 ~\mathrm{GeV}^2/c^4 \right) = 0.846^{+0.060+0.016}_{-0.054-0.014}$, which is $2.5\sigma$ from the SM prediction. The improvement in uncertainties has been compensated by a shift in the central value to leave the significance of the observation almost unchanged. Further updates are eagerly anticipated using the additional 4~fb$^{-1}$ of 13~TeV data for $R_K$ and all 6~fb$^{-1}$ data for the updates to $R_{K^*}$ to see if there will be any evidence of greater than $3\sigma$ departures from the SM.
It is also worth noting that very recent results by the  Belle Collaboration~\cite{belleRK} are consistent with both LHCb and the SM, albeit with larger error bars.

\begin{figure}[t]
\begin{center}
 \includegraphics[height=0.45\columnwidth]{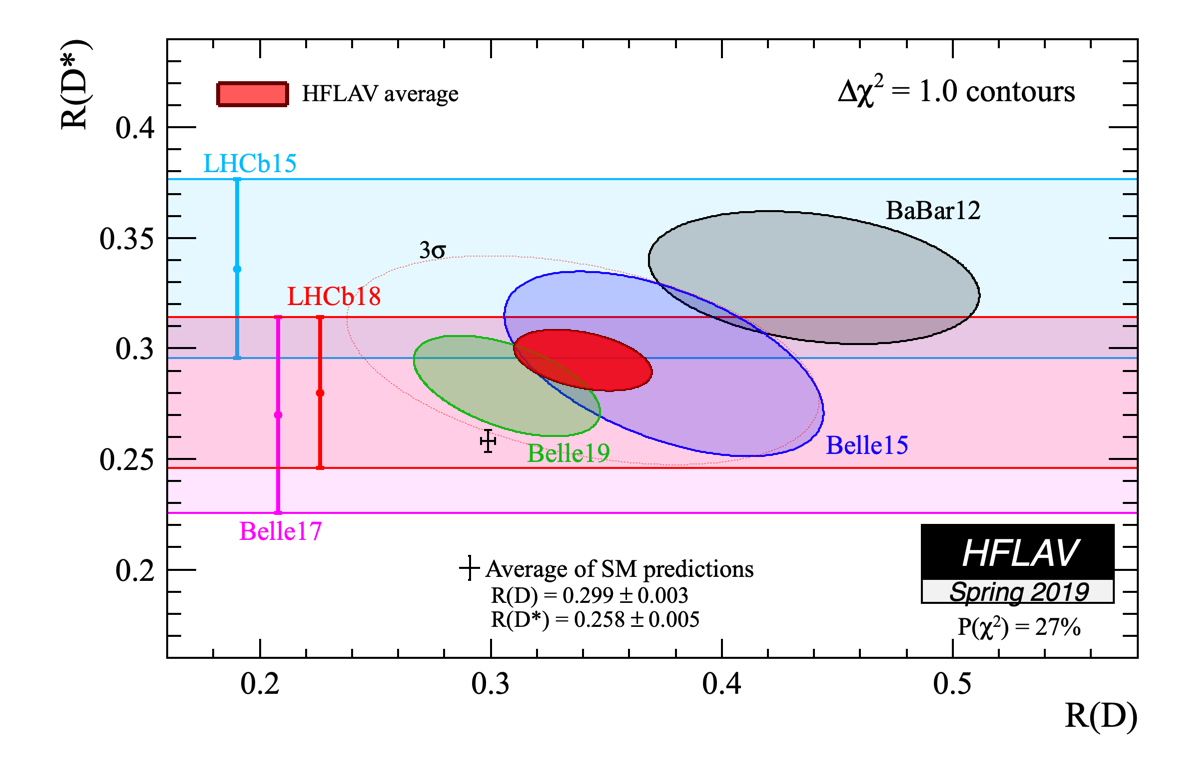}
\caption{Comparison of individual measurements, world average values and the theoretical predictions of $R(D)$ and R(D) \cite{hflav}.}\label{fig:RDstar}
\end{center}
\vspace{-1cm}
\end{figure}

The second set of measurements related to lepton universality are of 
$R\left(D^{(*)}\right) =\frac{\mathcal{B}\left(B\to D^{(*)}\tau^-\overline{\nu}_{\tau}\right)}{\mathcal{B}\left(B\to D^{(*)}\ell^-\overline{\nu}_{\ell}\right)}
$. Figure~\ref{fig:RDstar} shows the current world average compared to the theoretical prediction. The average disagrees with the SM by $3\sigma$, but the most recent measurement from the Belle Collaboration \cite{belle2019}, which uses a semileptonic tag, agrees with the SM prediction within uncertainties. Therefore, further measurements by Belle II and LHCb are required. It should be noted that these decays proceed at tree level, rather than the loop level of the electroweak penguins. Therefore, if there is new physics contributing to this decay the mass scale of the particles involved must be close to the electroweak scale $\mathcal{O}\left(1~{\mathrm TeV}\right)$, whereas the loop mediated processes are sensitive to much heavier mass scales as well.  

\begin{figure}[t]
\begin{center}
 \includegraphics[height=0.45\columnwidth]{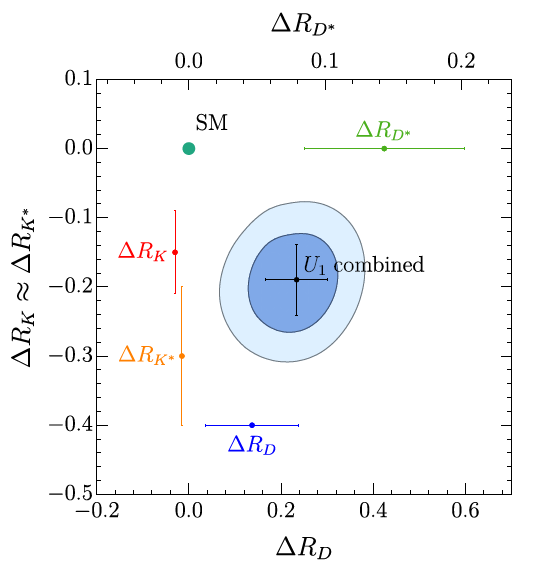}
\caption{Comparison of the measured discrepancies $\Delta R_{K^{(\ast)}}$ and $\Delta R(D^{(\ast)})$ with the SM and the predictions of a model of $U(1)$ vector leptoquarks \cite{leptoquarks}.}\label{fig:leptoquarks}
\end{center}
\end{figure}

The third anomalous measurement is in the differential  decay rate and angular distribution of the decay $B^{0}\to K^{\ast} \mu^{+}\mu^{-}$ \cite{LHCb_P5prime},
particularly the observable $P_{5}^{\prime}$ \cite{P5prime} where there is $3.4\sigma$ discrepancy with the SM prediction reported by the LHCb Collaboration. This anomaly appears more generally across the landscape of exclusive $B^0\to K^{(*)}\mu^+\mu^-$ and $B_s\to \phi\mu^+\mu^-$ observables~\cite{Descotes-Genon:2013wba,Descotes-Genon:2015uva}.
The LHCb analyses of $b\to s \mu^+\mu^-$ angular distributions are yet to be updated with the $6~\mathrm{fb}^{-1}$ of data collected at 13~TeV, and similar measurements have been reported by Belle~\cite{belle}, ATLAS~\cite{atlas} and CMS~\cite{cms}.
From the theory point of view, predictions for these observables require a handle on local and non-local form factors~\cite{Altmannshofer:2008dz,Bobeth:2017vxj}.
These can be addressed systematically at low dilepton invariant mass from light-cone sum rules with $B$-meson distribution amplitudes (DAs)~\cite{Khodjamirian:2006st,Khodjamirian:2010vf,Gubernari:2018wyi,Descotes-Genon:2019bud}. For that purpose, continuing efforts in the study of the $B$-meson DAs~\cite{Braun:2017liq,Beneke:2018wjp} look promising.

Model-independent analyses, with lepton universality conserving and violating Wilson coefficients, in which all the anomalies taken together in both $b\to s\ell^+\ell^-$ and $b\to c\tau^{-}\overline{\nu}_{\tau}$ decays, show evidence for physics beyond the SM at the level of $7\sigma$ \cite{NewPhysicsAnomalies}. Among the specific extensions of the SM, those that contain vector leptoquarks are most promising to describe all the data coherently. An example, from Ref.~\cite{leptoquarks}, is shown in Fig.~\ref{fig:leptoquarks}, where all the anomalies are described consistently with the introduction of a $U(1)$ vector leptoquark.   
   
Until there is undeniable statistical evidence for these discrepancies, there will be continued debate about whether these measurements are providing the first collider-based hints of beyond-the-SM physics. Fortunately, with the upgraded LHCb \cite{lhcb_upgrade} and Belle II experiments \cite{b2tip} about to collect unprecedentedly large samples of $b$ decay, there is great potential for discovery in the years ahead.
 
\section{Conclusion}
The study of heavy flavour physics is a vibrant topic with results probing the QCD, electroweak physics and searching for physics beyond the SM. Furthermore, we have only been able to highlight a handful of the many results presented during the working-group sessions due to concise nature of this summary. This vibrancy within the community will continue, as the particle physics community focusses more and more on precision measurements and searches for rare processes, given the absence of the direct evidence for new physics at the TeV scale from the results reported by ATLAS and CMS. The potential for discovery in flavour physics is great and we expect many exciting $c$, $b$ and $t$ decay results to be presented at future DIS workshops. 

\section{Acknowledgments}
We thank all the speakers in WG5 for their presentations and their contribution to the lively discussion. We thank the organizers for giving us this opportunity to convene the group and interact with the DIS community; this was a novel and educational experience for all three of us. We also thank them for a very well organized conference in a beautiful and gastronomically stimulating location.
V.B. would like to thank SERB (India) for providing financial support though an INSPIRE Faculty grant.
J.V. acknowledges funding from the European Union's Horizon 2020 research and innovation programme under the Marie Sklodowska-Curie grant agreement No 700525 `NIOBE', and support from SEJI/2018/033 (Generalitat Valenciana).

\end{document}